\documentstyle[12pt]{article}

\newcommand{\bt}{\beta}
\newcommand{\beg}{\begin{equation}}
\newcommand{\enq}{\end{equation}}
\newcommand{\sig}{\sigma}
\newcommand{\cst}{{\rm constant}}

\newcommand{\ap}{\alpha}
\newcommand{\cn}{\cal N}
\begin{document}

\title{Gap Formation Probability of the $\ap-$ Ensemble }
\author{Yang Chen and Kasper Juel Eriksen$^{\dag}$\\
        Department of Mathematics, Imperial College\\
        180 Queen's Gate, London SW7 2BZ, U. K.\\
        $^{\dag}$ {\O}rsted Laboratory,
        Niels Bohr Institute\\
        H.C.{\O}. Universitetsparken 5,
        2100 Kbh. {\O}, Denmark}
\date{\today}
\maketitle

\begin{abstract}
In this paper we employ the continuum approximation of Dyson to determine
the asymptotic gap formation probability in the spectrum of
$N\times N$ Hermitean matrices associated with orthogonal polynomials
that are solutions of indeterminate moment problems.
\end{abstract}

\newpage

\noindent
{\bf I Introduction}\\
Random matrix ensembles originally proposed by Wigner as a
phenomenological description of the statistical properties of
the energy spectra
\cite{Wigner},\cite{Porter} has recently seen applications in other
area of physics such as Quantum Chaos\cite{Giannoni}, transport in
disordered systems\cite{Stone} and 2-d quantum gravity\cite{Gross}.
{}From the random matrix point of view\cite{Mehta},\cite{Tracy1}, the
quantities of interest (in increasing order of refinement) are
the eigenvalue density, the density-density correlation function and
$E_{\bt}[n,J]$---the probability that an interval, $J,$
contains exactly $n$ eigenvalues. [This
interval can be generalized to the case where it is the union of several
mutually disjoint intervals\cite{Tracy2},\cite{Widom}].\\
This paper shall focus on $E_{\bt}[0,J],$ henceforth denoted as $E_{\bt}[J].$
The
probability that there is no
eigenvalue in a scaled interval $(-t,t)$, is
$\ln E_{\bt}(-t,t)\sim -\frac{\pi^2}{4}\bt t^2-(1-\bt/2)t $; first
determined by Dyson\cite{Dyson}, using the methods of classical
electrostatics, potential theory and thermodynamics, for random ensembles
with unitary $(\bt =2)$, orthogonal $(\bt=1)$ and symplectic $(\bt=4)$
symmetries. The missing constant and $O({\rm ln} t)$ terms were later found in
\cite{Widom1},\cite{Mehta1} and \cite{Dyson2}.\\
For Hermitean matrices
$(\bt=2)$ a fundamental result of Gaudin and Mehta \cite{Mehta} shows
that
\beg
E_2[J]=\det(1-K_J),
\enq
 where
$K_J$ is the kernel of an integral operator $K(x,y)$ over $J$;
\beg
K(x,y)={\rm e}^{-u(x)/2}{\rm e}^{-u(y)/2}\sum_{n=0}^{N-1}
p_n(x)p_n(y),
\enq
\beg
\int_{P}dx\;w(x)p_n(x)p_m(x)=\delta_{mn},\;\;w(x):={\rm e}^{-u(x)},
\enq
where $P$ is the interval available to the eigenvalues. Here $u(x)$ in the
Coulomb gas
interpretation is an external (confining) potential that holds
together a gas of repelling particles.
An alternative representation for the Fredholm determinant is
\beg
E_{\bt}[J]=\frac{Z(J^c)}{Z(J\cup J^c)}
=:{\rm e}^{-(F[J]-F[J\cup J^c])},
\enq
recognized as the ratio of two partition functions, where
\beg
Z[J]=\left(\prod_{a=1}^{N}\int_{J}dx_a\right){\rm e}^{-W}
\enq
and
\beg
W=-{\bt\sum_{0\leq a<b\leq N}\ln|x_a-x_b|+\sum_{a=0}^{N}u(x_a)}.
\enq
A Fredholm determinant of this kind also appears in context of the
physics of interacting bosons in one dimension\cite{Korepin} where
it gives the probability of emptyness formation of a segment of the real
line.\\
For sufficiently large $N$ it is expected that the discrete Coulomb
gas can be well approximated by a continuum Coulomb Fluid\cite{Dyson3} with
an equilibrium density $\sig(x)$ of charged particles
that satisfies a H\"uckel-like self consistent equation;
\beg
u(x)-\bt\int_{J}dy\sig(y)\ln|x-y|+\left(1-\frac{\bt}{2}\right)
\ln\left[\sig(x)\right]=A=\cst,
\enq
and gives the free energy, at equilibrium, with precisely $N$ charges
(eigenvalues) contained in an interval $J$;
\beg
F[J]=\frac{1}{2}AN+\frac{1}{2} \int_{J}dx u(x)\sig(x)-\frac{1}{2}
\left(1-\frac{\bt}{2}\right)\int_{J}dx\sig(x)\ln\left[\sig(x)\right],
\enq
subject to $\int_{J}dx\sig(x)=N.$
This approach has the advantage of being more general and has been shown
to supply accurate asymptotics of the level-spacing distribution for the
Laguerre ensemble at the hard edge\cite{Chen1}
for all $\bt > 0$. We expect
that the continuum Coulomb Fluid description will also be useful
in determining $E_{\bt}[J]$ of orthogonal
polynomial random matrix
ensembles associated with non-classical weight, $w(x)={\rm e}^{-u(x)},$
where a explicit form of the polynomials are not known.\\
This paper is organized as follows: In section {\bf II},
we shall introduce and discuss the properties of Hermitian
random ensembles (hence forth called the $\ap-$ensembles) with
$u(x)=x^{\ap}/\ap,(\;\ap>0,\;x\in(0,\infty))$. It will be shown in the
continuum approach the parameter $\ap$ plays the role of a coupling
constant with a critical value $(\ap_c),$ above which the density is
universal (i.e. $\ap$ independent) at both the hard and soft edges
\cite{Tracy3}. For $\ap>\ap_c$, the density at the soft edge
remains universal, however, an explicit $\ap$ dependence is found
at the hard edge.\\
By using a heuristic argument based on the leading asymptotics found by
Dyson\cite{Dyson} follow by an un-folding transformation of
the density, we obtained an asymptotic formula for, $E_2[J],$ the probability
that the interval $J$ contains no eigenvalue.
This is first used to test against known results \cite{Tracy3},\cite{Chen1}
and later
make predictions on the gap formation probability of the $\ap$ ensembles.\\
In section {\bf III} a screening theory of the Coulomb Fluid is presented
to justify the results obtained in section {\bf II}. Level
Spacing Distribution for $\ap=1/2$ is case presented in detail in section
{\bf IV}. This paper concludes with section {\bf V}, in which we
give $E_{2}[J]$ for certain random matrix ensemble arising from the
double scaling limit of two-dimensional quantum gravity.\\
\noindent
{\bf II Behaviour of density at edges}\\
As we shall be interested exclusively in $\bt=2$ case,
Eq.(7) becomes linear and reads;
\beg
u(x)-\bt\int_{0}^{b}dy\sig(y)\ln|x-y|=A,
\enq
where $b,$ the upper band edge, is related to the total number of particle
(i.e. eigenvalues) through $\int_{0}^{b}dy\sig(y)=N.$
By taking a derivative with respect to  $x$, Eq.(9) is
converted into an integral equation with Cauchy kernel;
\beg
u^{\prime}(x)=\bt{\rm P}\int_{0}^{b}dy\frac{\sig(y)}{x-y},
\enq
whose solution can be obtained by a standard technique\cite{Akhiezer1};
\beg
\sig(x)=\frac{1}{\bt \pi^2}{\sqrt {\frac{b-x}{x}}}{\rm P}\int_{0}^{b}
\frac{dy}{y-x}{\sqrt {\frac{y}{b-y}}}u^{\prime}(y).
\enq
Since
\beg
{\rm P}\int_{0}^{b}\frac{dy}{x-y}\frac{1}{{\sqrt {y(b-y)}}}=0,
\enq
the general solution of Eq.(10) is found by adding to Eq.(11)
the solution of the homogeneous equation, $\frac{\cst}{{\sqrt {x(b-x)}}}.$
However, it is clear that by including the homogeneous solution, the
equilibrium
free energy will be larger. The homogeneous solution will
not be included.
It can also be verified by direct computation that the density obtained
without the homogeneous solution compares well with
the density obtained from
orthogonal polynomials.\\
A generalization to the case of finitely many non-overlapping has been
studied\\
by Akhiezer\cite{Akhiezer1};
\beg
u^{\prime}(x)=\bt{\rm P}\int_{J}dy\frac{\sig(y)}{x-y},\;\;x\in J
\enq
where
\beg
J=\cup_{k=1}^{p}J_k,\;\;\; J_k=(a_k,b_k),\;\;\;a_k<b_k.
\enq
The solution to Eq.(13) is
\beg
\sig(x)=\frac{1}{\pi^2\bt}
{\sqrt {\prod_{k=1}^{p}\left(\frac{b_k-x}{x-a_k}\right)}}
{\rm P}\int_{J}\frac{dy}{y-x}{\sqrt
{\prod_{l=1}^{p}\left(\frac{b_l-y}{y-a_l}\right)}}
u^{\prime}(y),\;\;x\in J.
\enq
This integral representation can be useful in determining the level spacing
distribution in the multiple interval case. \\
If we now take
$u(x)=\frac{1}{\ap}\;x^{\ap},\ap>0$
then
$$
\sig(x)=\frac{1}{\pi^2\bt}{\sqrt {{b-x\over x}}}\int_{0}^{b}
\frac{dy}{y-x}{\sqrt {{y\over b-y}}}\;y^{\ap-1}
$$
\beg
=\frac{b^{\ap-1}}{\pi^2\bt}\frac{2\Gamma(1/2)\Gamma(\ap+1/2)}
{\Gamma(\ap)}{\sqrt {{b-x\over x}}}\;F(1-\ap,1,3/2,1-x/b),
\enq
where $F(a,b,c,x)=\;_2F_1(a,b;c;x)$ is the Hypergeometric function.
Note that for $\ap=1$, the result for the Laguerre Ensemble is recovered;
$\sig(x)=\frac{1}{\pi\bt}{\sqrt {{b-x\over x}}}.$
We have in the Hypergeometric function the following parameters,
$a=1-\ap,\;\;b=1,\;\;c=3/2,\quad {\rm and}\quad  a+b-c=\frac{1}{2}-\ap.$
 From this the behaviour of the density at the the
hard $(x\sim 0)$ and
soft $(x\sim b)$ edges\cite{Tracy3} can be determined.
Consider $\ap>1/2,$ and $x\ll b$, we find from the Gauss summation formula,
$$F(1-\ap,1,3/2,1)=\frac{\Gamma(3/2)\Gamma(\ap-1/2)}
{\Gamma(1/2)\Gamma(\ap+1/2)},$$
\beg
\sig(x)\sim \frac{1}{\pi^{3/2}\bt}\frac{\Gamma(\ap-1/2)}{\Gamma(\ap)}
b^{\ap-1}\;{\sqrt {{b\over x}}},\;\;x\ll b,
\enq
The density has a square root singularity at $x=0,$ independent of $\ap$
and is therefore universal.
Near the soft edge, $x\sim b,$
\beg
\sig(x)\sim \frac{1}{\pi^{3/2}\bt}\frac{\Gamma(\ap+1/2)}{\Gamma(\ap)}
b^{\ap-1}{\sqrt {{b-x\over b}}},\;\;b-x\ll b,
\enq
which shows that the density is universal near $b$.
However, for $|\ap|<1/2$, a completely different behaviour
at the hard edge occurs, this can be seen as follows:
A translation,$x\rightarrow 1-x$ gives,
$$
F(1-\ap,1,3/2,1-x)=\frac{\Gamma(3/2)\Gamma(\ap-1/2)}
{\Gamma(\ap+1/2)\Gamma(1/2)}F(1-\ap,1,3/2-\ap,x)
$$
\beg
+
x^{\ap-1/2}\frac{\Gamma(3/2)\Gamma(1/2-\ap)}{\Gamma(1-\ap)}
F(\ap+1/2,1/2,\ap+1/2,x).\enq
In view of the combination of the parameters $(a,b,c)$,
the second term of the right hand side of Eq.(19)
dominates as $x\rightarrow 0$.
Putting $x=0$ in the second Hypergeometric function;
we find $F(a,b,c,0)=1,\;\; 0<a+b-c<1.$ Therefore,
\beg
\sig(x)\sim \frac{\tan\pi\ap}{\pi\bt}\;\;\frac{1}{x^{1-\ap}},\;\; x\ll b.
\enq
Eq.(20) has two features that are worth noting;
1. $\sig(x)$ near the hard edge is independent of the macroscopic parameter
$b$ [which increases with $N$]. This type of behaviour at the hard edge
is also be observed in the unitary $q-$ Laguerre Ensemble with
$u(x;q)=\sum_{n=0}^{\infty}\ln[1+(1-q)xq^n],\;0<q<1\;x\geq 0;$
$$\sig_N(0)=\frac{1-q^N}{\ln \left(\frac{1}{q}\right)},$$ which
rapidly becomes $N$ independent as $N$ increases\cite{Chen5}, 2. The
hard edge density has a explicit $\ap$ dependence.\\
We remark here that the kernel in the unitary $q-$ Laguerre Ensemble
reads,
$$
K(x,y)
$$
\beg
=\frac{\cst}{x-y}\left[\left(\frac{x}{y}\right)^{1/4}
\sin\left(\frac{\pi}{2\gamma}\ln x\right)
\cos\left(\frac{\pi}{2\gamma}\ln y\right)-\left(\frac{y}{x}\right)^{1/4}
\sin\left(\frac{\pi}{2\gamma}\ln y\right)\cos\left(\frac{\pi}{2\bt}
\ln x\right)\right],
\enq
in which $\gamma:=\ln(1/q)\gg 1$ and $N\rightarrow \infty$\cite{Chen6}.
Therefore the so-called universal density-density correlation function
at the hard edge $<\rho(x)\rho(y)>-<\rho(x)><\rho(y)>=\delta(x-y)<\rho(x)>
-[K(x,y)]^2=(xy)^{-1/2}(x^{1/2}+y^{1/2})/(x-y)^2$ \cite{Bee} is seen to be
violated. This is due to the weakly confining nature of the
$q-$ Laguerre potential; $u(x;q)\sim [\ln x]^2,\;\;x\rightarrow \infty,$
producing a heavily depleted density at the origin. We expect in the
$\ap-$ ensemble a similar phenomena to occur for $\ap<1/2$. In principle,
the gap formation probability of the unitary $\ap-$ ensemble at the
hard edge can be computed directly from the Fredholm determinant in Eq.(1),
however, the associated orthogonal polynomials being related to
an indeterminate moment problem\cite{Akhiezer2,Mourad} is not
known, unlike the unitary $q-$ Laguerre Ensemble\cite{Chen10}.
A direct evaluation
of the Fredholm determinant would be difficult. To extract the leading
term in $-\ln E_2(0,s)$ we shall adopt the Coulomb Fluid approximation
mentioned in section {\bf I}.\\
Before we give results on the $\ap-$ ensemble, it is interesting to note
that for $\ap=1/2,$
\beg
\sig(x)=\frac{1}{\pi^2\bt}\;\frac{1}{{\sqrt x}}
\ln\left[{1+{\sqrt {1-x/b}}\over {1-{\sqrt {1-x/b}}}}\right],\;\;b=\pi^2N^2
\enq
where we have made use of the identity
$$F(1/2,1,3/2,x)=\frac{1}{2{\sqrt x}}
\ln\left[{1+{\sqrt x}\over 1-{\sqrt x}}\right],$$
to arrive at Eq.(22).
\par\noindent
Thus for $x\ll b$
\beg
\sig(x)\sim \frac{1}{\pi^2\bt}\;\frac{\ln(4b/x)}{{\sqrt x}}+
{\rm o}({\sqrt x}),
\enq
while for $x\sim b$
$$\sig(x)\sim \frac{1}{\pi^2\bt}\left[(1-x/b)^{1/2}+\frac{5}{3}(1-x/b)^{3/2}+
\cdots\right],$$
remains to be universal. We may therefore interpret $\ap=1/2$ to be
the critical point at which there is a logarithmic correction to the
density.\\
In order to gain insight into the possible leading terms in
the gap formation probability, we examine the results first
obtained by Dyson for the Circular ensemble\cite{Dyson} where by
construction the density is uniform over the circle, $N/2\pi.$ It was
shown that
$-\ln E_2(0,s)\sim \cst\; s^2,\;\;s\gg 1.$ Here $s$ being expressed
in terms of the average distance between the charges is dimensionless. The
leading term is simply proportional to the square of the number of particles
excluded in an interval of length $s.$ Using the unfolding
transformation in the density, we can ${\it always}$ define a new
variable $y$, such that $\sig(x)dx=dy$, or equivalently
$\frac{dy}{dx}=\sig(x)$. The density in $y$ is unity. At this point,
we may make use of Dyson's asymptotic and conclude that the leading term
is proportional
to the square of the number of particles excluded in that interval.
The probability that the interval $J$ contains no eigenvalues is
\beg
-\ln E_{\bt}[J]\sim\left[\int_{J}dx\sig(x)\right]^2.
\enq
Based on Eq.(24), the probability that an interval of length $a$ (measured
from the origin) contains no eigenvalue is,
$$E_{\bt}[a]\sim {\rm e}^{-\left[{\cn}(a)\right]^2},$$
where
$${\cn}(a):=\int_{0}^{a}dx\sig(x),$$
is the number of charges excluded in $(0,a).$
Taking $\sig(x)=\frac{1}{\pi\bt}{\sqrt {\frac{b-x}{x}}},$
we find
\beg
\int_{0}^adx\sig(x)=\frac{b}{\pi\bt}\int_{0}^{a/b}
{\sqrt {\frac{1-x}{x}}}\sim \frac{2}{\pi\bt}\left[{\sqrt {ab}}-
\frac{1}{6}\frac{(ab)^{3/2}}{b^2}+\cdots\right].
\enq
In the limit $b\rightarrow \infty$ (or $N\rightarrow \infty$, since
$b\propto N$) and $a\rightarrow 0$ such that
the combination $ab=s$ is finite, gives
\beg
-\ln E_{\bt}(0,s)\sim s,
\enq
agrees with the know leading behaviour\cite{Tracy3},\cite{Forrester},
\cite{Chen1}. By
taking $J=(a,b),\; a<<b$ and scaling $b-a$ with respect to the true soft
edge density $\sig_N(N)\sim N^{-1/3}\sim b^{-1/3}$, we find,
\beg
-\ln E_{\bt}(0,s)\sim s^3,\;\;\;{\rm where}\;
s:=\left(\frac{b-a}{b^{1/3}}\right),
\enq
which agrees with
that found in\cite{Tracy3},\cite{Chen7}.
Observe that although the density in the continuum approach
vanishes at $b,$ the correct dependence
at the soft edge, i.e., $\sig(b)\sim b^{-1/3},$ is found\cite{Forrester}.
\par\noindent
For the $\ap-$ ensembles,
\beg
-\ln E_{\bt}^{(\ap)}(0,a)\sim a^{2\ap},\;\;\;0<\ap<1/2,
\enq
while $\ap=1/2$,
\beg
{\sqrt {-\ln E_{\bt}^{(1/2)}(0,a)}}
\sim \int_{0}^{a}dx \frac{\ln(2\bt b/x)}{{\sqrt x}}
\sim {\sqrt a}\ln\left(\frac{2\bt b}{a}\right).
\enq
\noindent
{\bf II Screening theory}\par\noindent
In this section we shall consider the unitary case $(\bt =2)$. With
a slight change of notation, the free energy of $N$ particles restricted
in an interval $I=(0,b)$ under the influence of an external potential
$u(x)/\bt $ is
\beg
F[\sig, I, N]=\int_{I}dx\,u(x)\,\sig(x,I,N)
-\int_{I}dx\int_{I}dy\sig(x,I,N)\ln|x-y|\sig(y,I,N),
\enq
where $\sig(x,I,N)$ is the equilibrium density that minimizes
$F[\sig,I,N]$,
subject to
$$\int_{I}\sig(x)dx=N.$$\par\noindent
According to Eq.(4) the quantity of interest is the change in free energy,
i.e. the free energy when the charges are excluded from the sub-interval
$J=(0,a),\;a<b$ of $I$ minus the free energy when the  charges are in $I$:
\beg
\Delta F[J,N]=F[J^c,N]-F[I,N],
\enq
in the limit $N\rightarrow \infty$. Here the interval $J^c$ is the
the interval where the charges are distributed, when there are no
charges in $J$. For later use we introduce,
\beg
\Delta\sig(x,J,N):=\sig(x,J^c,N)-\sig(x,I,N),
\enq
and require that all the $\sig'$s are zero outside their respective
supports. The support of $\Delta\sig(x,J,N)$ is $L=I\cup J^c.$\\
Using the definitions of $\Delta F$ and $\Delta\sig$ in Eqs.(30) and (31), a
simple calculation gives,
$$
\Delta F[J,N]=\int_Ldx \left[u(x)-2\int_Idy
\ln|x-y|\sig(y,I,N)\right]\Delta\sig(x,J,N)$$
\beg
-\frac{1}{2}\int_{L}dx\int_{L}dy\Delta\sig(x,J,N)\ln|x-y|
\Delta\sig(y,J,N).
\enq
Since $\sig(x,I,N)$ is the density that minimizes the
free energy; $[\cdots]$, in Eq.(33) is equal to the
chemical potential, $\mu(I,N),$ for $x\in I$ and is a function of
of $x$, for $x\in L-I =J^c -I.$
Charge neutrality states that $\int_L dx \,\Delta\sig(x,J,N) = 0.$
Therefore the first term in Eq.(33) can be rewritten as
$$
\int_Ldx \left[u(x)-2\int_Idy
\ln|x-y|\sig(y,I,N)\right]\Delta\sig(x,J,N)
$$
$$
= \int_Ldx \left[u(x)-2\int_Idy
\ln|x-y|\sig(y,I,N)-\mu(I,N) \right]\Delta\sig(x,J,N)
$$
$$
= \int_{L-I}dx \left[u(x)-2\int_Idy
\ln|x-y|\sig(y,I,N)-\mu(I,N) \right]\Delta\sig(x,J,N).
$$
The term $[\cdots]$ under $\int_{L-I}dx$ is of the same order of magnitude
as the change in the chemical potential when the particles are
excluded from $J$: $\Delta\mu(J,N) = \mu(J^c,N) - \mu(I,N)$.
$\Delta\mu(J,N)$ decreases as $N$ grows and in the limit,
$N\rightarrow \infty,$ it is proportional to the number of particles
excluded from $J$. Therefore the first term in Eq.(33) is
approximately $\Delta\mu(J,N)$ times the number of particles in
$J^c-I$, which is the number of particles that spilled over
when {\it all} particles are excluded from $J$. In the bulk and at
the hard edge this number goes to zero when
$N\rightarrow \infty$ and therefore the first term in Eq.(33)
can be neglected.
At the soft edge the first term contributes a term quadratic
in the number of particles excluded from $J$. This is
of the expected form from the unfolding transformation and will not affect
our final result.
We therefore have,
\beg
\Delta F[J,N]=-\frac{1}{2}\int_{L}dx\int_{L}dy\Delta\sig(x,J,N)
\ln|x-y|\Delta\sig(y,J,N),
\enq
here $\Delta\sig(x,J,N)$ is simply $-\sig(x,I,N)$ for $x\in J.$ For
$x\not\in J$, the expression is more complicated. From now on
we shall work exclusively with rescaled quantities such that lengths
are measured in the units of $|J|(=a)$. We also translate the origin
of the real axis to the left end of $J.$
With the scaling, $x\rightarrow ax,$ and $y\rightarrow ay,$ and the
introduction of $\rho(x)$ and $\lambda(x),$
\beg
\rho(x):=\rho(x,J,N):=-a\sig(ax),\;\;x\in (0,1)
\enq
\beg
\lambda(x):=\lambda(x,J,N)=a\Delta(ax,J,N),\;\;\;x\in [1,b/a).
\enq
The change in free energy when expressed in terms of
scaled quantities reads,
$$
\Delta F[J,N]=-\frac{a^2}{2}\int_Ldx\int_Ldy
\Delta\sig(ax,a)\ln|ax-ay|\Delta\sig(ay,a)
$$
$$=-\frac{\ln a}{2}\left[\int_Ldx\Delta\sig(x,J)\right]^2
-\frac{a^2}{2}\int_{L/a}dx\int_{L/a}dy\Delta\sig(ax,a)\ln|x-y|
\Delta(ay,a)
$$
$$
=\int_{0}^{1}dx\int_{J^c/a}dy\rho(x)\ln|x-y|
\lambda(y)
$$
\beg
-\frac{1}{2}\int_{J^c/a}dx\int_{J^c/a}dy
\lambda(x)\ln|x-y|\lambda(y)
-\frac{1}{2}\int_{0}^{1}dx\int_{0}^{1}dy\rho(x)\ln|x-y|
\rho(y).
\enq
Note that the first integral in the second equality of Eq. (37) vanishes due to
charge neutrality,$\int_Ldx\Delta\sig(x,J)=0.$
The problem of finding $\Delta\sig(x,J,N)$ in the limit
$N\rightarrow \infty$ can be reformulated in physical terms.
The interval $I$ is force (or field) free
since $I$ is a conducting region. Outside $I$ we have
a potential that grows when $I$ tends to infinity. We introduce
negative charges in $J$ with density $-\sig(x,I,N)$ and precisely the same
numbers of positive charges outside $J$.[The is the charge neutrality
condition stated above.] The charges redistribute themselves in
such a way that
the interval $L-J$ is force free again. The charges in $J$
are screened by those outside $J$. This screening process is not
as efficient as in three dimension since the charges are only allowed
to be built up along a line while the force extends out in a
plane. $\lambda(x,J)$ is then determined by the condition that
the total force in $J^c/a$ is zero.\\
The force produced by the charges in $J$ is
\beg
f(x)=\int_{0}^{1}dy\frac{\rho(y)}{x-y}.
\enq
The force produced by the charges in $J^c$ is
\beg
f_{J^c}(x)=\int_{J^c/a}dy\frac{\lambda(y)}{x-y}.
\enq
Therefore the total force is
\beg
f_{J^c}(x)+f(x)+f_I(x)=0,\;\;x\in J^c/a,
\enq
where
$f_I$ is the combined force in $I,$
produced by the external potential $u(x)/2$ and the charge
distribution $\Delta\sig(x,I,N)$.\\
Two constraints are required to solve
Eq.(40): 1. The number of charges in $J^c/a$ equals that
in $J/a$, 2. $\Delta\sig(x,J^c)$ is positive semi-definite for
$x\in J^c,$ and minimizes the free energy.
We shall also require that at an end point of $J^c/a$
the density vanishes. The reason is
simply that the charges would otherwise distribute in a larger
interval. This argument can be generalized to the case
where $J^c=\cup_{k=1}^{p}I_k,$ with the additional requirement that
the chemical potentials to be the same in each intervals $I_k$.
The above formulation can be applied to $\lambda(x)$.
In the above scheme we make the following
approximation: If $J$ is not at the end of $I$ where new charges
gather, the total potential can be approximated by a constant
all the way up to infinity and not only inside $I$. Observe that
in this approximation $\lambda$ is supported everywhere on the
real axis outside $J$; except in the hard edge case where it is only
a half line. Call this interval $P$. In the rescaled units
we have as an example in the hard edge problem, $P=(1,\infty)$, while
in the bulk problem, $P=(-\infty,0)\cup(1,\infty).$
In terms of the interval $P$, the change in free energy reads,
$$
\Delta F[J,N]=+\int_{0}^{1}dx\int_{P}dy\rho(x)\ln|x-y|\lambda(y)
$$
\beg
-\frac{1}{2}\int_{P}dx\int_{P}dy \lambda(x)\ln|x-y|\lambda(x)
-\frac{1}{2}\int_{0}^{1}dx\int_{0}^{1}dy\rho(x)\ln|x-y|\rho(y).
\enq
To proceed further, we can express $\lambda$ in terms of $\rho$ by
the use of the condition that the total force is $0$ in $P$. As above
we find that the total force is
\beg
\int_P dx\frac{\lambda(x)}{x-y}+f(y)=0.
\enq
The solution of Eq.(42) depends only on $P$ and
$\lambda|_{x\in \partial P}$. For, $P=(1,\infty)$,
the solution is
\beg
\lambda(x)=\frac{1}{\pi^2}\frac{1}{{\sqrt {x-1}}}\int_{1}^{\infty}
\frac{dy}{x-y}{\sqrt {y-1}}f(y).
\enq
The most important feature about this solution is that
$\lambda$ is a linear functional of $f$ and therefore also of
$\rho$, (see Eq.(40)), which does not depend on $a$ and $N.$
To proceed further, we now develop $f$ in a ``multipole'' (or moment)
expansion;
$$
f(x)=\int_{0}^{1}dy\frac{\rho(y)}{x-y},\;\;x\in (1,\infty)
$$
\beg
=\sum_{m=0}^{\infty}\frac{1}{x^{m+1}}\int_{0}^{1}dyy^m\rho(y).
\enq
The moments
\beg
A_m:=\int_{0}^{1}dxx^m\rho(x),
\enq
will play an important role in a later development. Since
$\lambda$ is a linear functional of $f$ and $f$ is a linear
functional of $A_m'$s, we conclude that $\lambda$ is a linear functional of
the $A_m'$s. By expanding ${\sqrt x}\rho(x),\;\;x\in [0,1]$, in terms
of Legendre polynomials $P_n(2x-1)$,
\beg
{\sqrt x}\rho(x)=\sum_{m=0}^{\infty}r_mP_m(2x-1),\;\;x\in [0,1]
\enq
we find also that ${\sqrt x}\rho(x)$ is a linear
functional of the moments $A_{m+\frac{1}{2}}.$
In summary, the densities can be expressed as linear functional
of $A_m$ and $A_{m+\frac{1}{2}}.$ Since the change in the free energy
is quadratic in the densities, it can be expressed
as a sum quadratic in the $A_m'$s. A simple estimate gives,
\beg
|A_m|\leq |A_0|,
\enq
from which we deduce that the change in free energy is
bounded from above by $A^2_0$. This we recognized as the square of the
number of charges excluded in $(0,a).$
Precisely the same analysis can be carried through in the bulk
scaling case.\\
The upshot of the above analysis is that $\Delta F$
is quadratic in the moments of the local densities, when
we adopt the Coulomb fluid approximation. From this we may infer two
consequences: 1. The probability of gap formation depends only
on the local densities. In the bulk scaling case the density
is always a constant and the
asymptotics level spacing distribution found by Dyson is reproduced,
2. Higher moments are nominally linear
functions of the number of excluded charges [a logarithmic
dependence is in general missed in the continuum approach,
because $\Delta F(J)$ disappears when the length of
$J$ tends to zero.].
Therefore, $\Delta F$ for large spacing is proportional
to the square of the number of excluded charges in the appropriate
interval. On the other hand it is shown that\cite{Chen1} the
gap formation distribution of the Laguerre Ensemble at the hard edge,
the large interval formula involves quadratic, linear and a logarithm
of the number of excluded charges ($\propto \sqrt s$). The breakdown
of the present theory can be traced back to the fact that the density
of the Laguerre Ensemble contains a point charge at
the origin and therefore can not be adequately described by its
moments. We may be conclude that leading term of $-\ln E_{\bt}[J],$
is proportional to ${\cn}^2[J],$ where ${\cn}[J]$ is the
number of charges excluded in $J.$ This justifies the
un-folding transformation discussed in section {\bf II}.\\
We now proceed to express the change in free energy in terms of
$A_m'$s; substitute Eq.(43) into Eq.(42), we have
$$
\lambda(x)=-\frac{1}{\pi^2}\frac{1}{{\sqrt {x-1}}}
\int_{1}^{\infty}\frac{dy}{x-y}{\sqrt {y-1}}\sum_{m=0}^{\infty}
\frac{2A_m}{(2y-1)^{m+1}}
$$
$$
=\sum_{m=0}^{\infty}\frac{2A_m}{\pi^2}
\frac{B(1/2,m+1/2)}{(2x-1)}\frac{F(-m,1,1/2,1/(2x-1))}{{\sqrt {2x-2}}}
$$
$$
=\sum_{m=0}^{\infty}\frac{2A_m}{\pi}\frac{1}{(2x-1){\sqrt {2x-2}}}
P_m^{(-1/2,1/2 -m)}\left(1-\frac{2}{2x-1}\right)
$$
\beg
=\sum_{0}^{\infty}\frac{2A_m}{m!\pi}
\frac{\left[1-1/(2x-1)\right]^{m-1}}{(2x-1)^2}
\frac{d^m}{dz^m}\left[{\sqrt {1-z}}z^{m-1/2}\right]\bigg|_{z=1/(2x-1)},
\enq
where $P^{\mu,\nu}_n(x)$ are the Jacobi polynomials.\\
With this expression for $\lambda(x)$ the second integral in the third
equality of Eq.(37) reads;
\beg
\frac{1}{2}\int_{1}^{\infty}dx\int_{1}^{\infty}dy\lambda(x)\ln|x-y|
\lambda(y)=\sum_{m=0}^{\infty}\sum_{n=0}^{\infty}A_mA_nL_{mn},\;\;
\enq
where
the matrix elements $L_{mn}=L_{nm}$ are
\beg
L_{mn}=\frac{1}{2\pi^2}\int_{0}^{1}ds\int_{0}^{1}dt
{\sqrt {\frac{t}{1-t}}}P_m^{(-1/2,1/2-m)}(1-2t)
\ln\bigg|\frac{1}{2t}-\frac{1}{2s}\bigg|{\sqrt {\frac{s}{1-s}}}
P_n^{(-1/2,1/2-n)}(1-2s).
\enq
The third integral, with the aid of Eq.(45), reads
\beg
\frac{1}{2}\int_{0}^{1}dx\int_{0}^{1}dy\rho(x)\ln|x-y|\rho(y)
=\sum_{m=0}^{\infty}\sum_{n=0}^{\infty}r_mr_nR_{mn},
\enq
where
\beg
R_{mn}:=\frac{1}{2}\int_{0}^{1}dx\int_{0}^{1}dy\ln|x-y|
\frac{P_m(2x-1)}{{\sqrt x}}\frac{P_n(2y-1)}{{\sqrt y}},
\enq
The first integral is,
\beg
\int_{0}^{1}dx\int_{1}^{\infty}dy\rho(x)\ln|x-y|\lambda(y)
=\sum_{m=0}^{\infty}\sum_{n=0}^{\infty}r_mA_nU_{mn},
\enq
where
\beg
U_{mn}:=\int_{0}^{1}ds\int_{0}^{1}dt\frac{P_m(2t-1)}{\sqrt t}
{\sqrt {\frac{s}{1-s}}}P_n^{(-1/2,1/2-n)}(1-2s)\ln\left[\frac{1}{2}
\left(\frac{1}{s}-(2t-1)\right)\right].\enq
As an example we determine the leading term of $-\ln E_{\bt}[J]$
of the $\frac{1}{2}-$ ensemble. The density in this case
is given by Eq.(23) with $\bt=2$.
We find the number of excluded charges in the
interval $(0,a),$
\beg
N(a)=\int_{0}^{a}dx\sig(x)\sim \frac{2}{\pi^2}{\sqrt a}\ln N,\;\;
N\rightarrow \infty.
\enq
In the limit $N\rightarrow \infty$ and $a\rightarrow 0$ such that
${\sqrt a}\ln N$ is finite, we find according
to the screening theory that
\beg
\Delta F\sim s, \;\;N\rightarrow \infty,\;\;s:=a\left[\ln N\right]^2
\;\;{\rm finite},
\enq
which agrees with the results obtained from the un-folding transformation
argument.
\par\noindent
{\bf IV Solution via the continuum approach for} $\ap=1/2$
\par\noindent
For $\ap=1/2$ the solution of the integral equation reads
$$
\sig(x)=\frac{1}{\pi^2\bt}{\sqrt {\frac{b-x}{x-a}}}\int_{a}^{b}
\frac{dy}{y-x}{\sqrt {\frac{y-a}{b-y}}\frac{1}{{\sqrt y}}}
$$
\beg
=\frac{2}{\pi^2\bt{\sqrt b}}{\sqrt {\frac{b-x}{x-a}}}\left(K(p)
+\frac{x-a}{b-x}\Pi\left(\frac{b-a}{b-x},p\right)\right),\;\;\;x\in (a,b),
\enq
where
$$K(p):=\int_{0}^{1}\frac{dy}{{\sqrt {(1-y^2)(1-p^2y^2)}}},
\;\;\;p^2:=\frac{b-a}{b}$$
is the complete elliptic function of the first kind, and
$$\Pi(n,p):=\int_{0}^{1}\frac{dy}{1-ny^2}
\frac{1}{{\sqrt {(1-y^2)(1-p^2y^2)}}},\;\;\;n:=\frac{b-a}{b-x},$$
is the complete elliptic function of the third kind.
Note that the $\sig(x)$ vanishes at $x=b$ and is
positive for $x\in (a,b).$
The upper band edge $b$ is related to $a$ and $N$ via the normalization
condition; $\int_{a}^{b}dx\sig(x)=N,$ as
\beg
N=\frac{2}{\pi\bt}{\sqrt a}\left(K(p)-D(p)\right),
\enq
where
$$
D(x):=\frac{1}{2}\int_{0}^{1}dt{\sqrt {\frac{t}{(1-t)(1-tx^2)}}}=
\frac{\pi}{4}F(1/2,3/2,2,x^2)$$
is an elliptic function. The derivation of Eq. (58) is
placed in Appendix A. In order to determine the
free energy in the interval $(a,b),$ we require the chemical potential
$A$ and the interaction energy $\frac{1}{2}\int_{a}^{b}dx\sig(x).$
The chemical potential reads
\beg
A=2{\sqrt b}-\bt\int_{a}^ {b}dx\sig(x)\ln|b-x|
=\bt N\left(2\ln 2-\ln b-\ln p^2\right)+\frac{4}{\pi}{\sqrt b}p^2E(p),
\enq
where a derivation of the above is placed in appendix B.
The interaction energy reads
\beg
\frac{1}{2}\int_{a}^{b}dxu(x)\sig(x)=\frac{2}{\pi^2\bt}(b-a)+
\frac{2}{\pi^2\bt}(b-a)\int_{0}^{a/b}dt\left[D\left({\sqrt {1-t}}\right)
\right]^2.
\enq
The derivation of Eq.(60) is placed in appendix C.\\
The free energy for $N$ charges residing in $(a,b)$ is recapitulated
as follows:
\beg
F(a,b)=\frac{\bt}{2}\left[2\ln 2-\ln(b-a)\right]+
\frac{2N{\sqrt b}p^2}{\pi}E(p)+\frac{2}{\bt\pi^2}\left[b-a\right]
\left[1+\int_{0}^{a/b}dtD^2\left({\sqrt {1-t}}\right)\right],
\enq
where
$$N=\frac{2{\sqrt b}}{\bt\pi}\left[E(p)-\frac{a}{b}K(p)\right].$$
We now compute the change in the free energy, by expressing $b$ as function
of $N$ and $a$ in the limits $N\rightarrow \infty$ and $a\rightarrow 0,$
after some straightforward but tedious calculations (not reproduced here),
we find
$$\Delta F\sim \frac{1}{\pi^2}\left[{\sqrt a}\ln N\right]^2
$$
which substantiates the result obtained from the screening theory.
\\
\noindent
{\bf V Conclusion}
\par\noindent
We have shown that applying the screening theory to the Coulomb fluid
model of random matrix ensemble, gap formation distribution,
which is the probability that there are no eigenvalue in a certain
interval,$J$, of the spectrum is, in the limit of large interval
$$
E_{\bt}[J]\sim {\rm e}^{-{\cn}^2[J]},
$$
where
$${\cn}[J]=\int_{J}dx\sig(x)={\rm number\;of\;eigenvalues\;excluded\;in\;}J.$$
This result shows that the determination of the
leading term in $-\ln E_{\bt}[J],$ requires only the integral of the
local density $\sig(x),$
which enables us to make prediction on the gap formation probability of
the $\ap-$ ensemble at the hard edge.\\
It is shown that for
$\ap>1/2,$ the leading term in the asymptotic expansion is independent
of $\ap$ with the scaling variable, $s\propto Na$. For $\ap=1/2$,
due to the logarithmic correction of the hard edge density
the scaling variable, $s\propto \left[\ln N\right]^2a.$
However, for $\ap<1/2$, the hard edge
density acquires $\ap$ dependence, and gives $s\propto N^0a.$
Universality is violated for $\ap<1/2$. Similar behaviour is found
in the $q-$Hermite
\cite{Chen1} and $q-$Laguerre ensembles\cite{Chen5}.\\
The advantage of this approach is that it by-passes the requirement
of the knowledge of the asymptotic expansion of non-classical orthogonal
polynomials, especially those associated with the indeterminate
classical moment problems.\\
As a side product of the screening theory, we make prediction
on the level spacing distribution in the soft edge
of Hermitian random matrix models that arises in double-scaling limit of the
two-dimensional quantum gravity. For example, taking
$u(x)=N(x+\frac{g}{4N}x^4),$ it can be shown that\cite{BIPZ}
tuning $g$ to $g_c\;(<0),$ the critical point density reads,
$$
\sig_c(x)=\frac{1}{24\pi}(8-x^2)^{3/2},
$$
where now $\int_{-{\sqrt 8}}^{\sqrt 8}dx\sig(x)=1.$
Note that in this formulation,
$$
\frac{F}{N^2}=-\frac{1}{2}\int_{-b}^{b}dx\int_{-b}^{b}dy\sig(x)\ln|x-y|\sig(y)
+\int_{-b}^{b}dx\sig(x)\left(x+\frac{g}{4N}x^4\right).
$$
Using our theory, a simple calculation gives,
$$
-\ln E_2(s)\sim N^2\left[\int_{a}^{\sqrt 8}dx\sig_c(x)\right]^2
\sim s^5
$$
where the scaling variable is $s:=N^{2/5}({\sqrt 8}-a),$ in
the limit $N\rightarrow \infty,$ $a\rightarrow {\sqrt 8}$ such
that $s$ is finite. The detail exposition for this and other
multi-critical point models can be found in\cite{Chen8}.
Work is underway to use this approach to determine, $E_{\bt}(s),$ of the
$q-$Laguerre ensembles.

\newpage
\noindent
{\bf Appendix A The normalization integral}
\par\noindent
By definition,
$$N=\int_{a}^{b}dx\sig(x)
=\frac{1}{\pi^2\bt}\frac{b-a}{{\sqrt b}}\int_{0}^{1}dy
{\sqrt {\frac{1-y}{y(1-p^2y)}}}\int_{0}^{1}\frac{dx}{x-y}
{\sqrt {\frac{x}{1-x}}}
$$
$$
=\frac{1}{\pi\bt}\frac{b-a}{{\sqrt b}}\int_{0}^{1}
dy{\sqrt {\frac{1-y}{y(1-p^2y)}}}=\frac{1}{\pi\bt}\frac{b-a}{{\sqrt b}}
B(1/2,3/2)F(1/2,1/2,2,p^2)
$$
$$
=\frac{2}{\pi\bt}{\sqrt b}p^2\left[K(p)-D(p)\right],\eqno(A.1)
$$
where we have made of
$$
F(1/2,1/2,2,p^2)=\frac{4}{\pi}\left[K(p)-D(p)\right].\eqno(A.2)$$
\newpage
\noindent
{\bf Appendix B Evaluation of the chemical potential}
\par\noindent
We have
$$
A=2{\sqrt b}-\bt N\ln(b-a)+I(a,b),\eqno(B.1)
$$
where
$$
I(a,b):=-\bt(b-a)\int_{0}^{1}dx\ln x\sig(b-x(b-a))
$$
$$
=-\frac{2{\sqrt b}}{\pi^2}p^2\int_{0}^{1}\frac{dy}{2{\sqrt y}}
{\sqrt {\frac{1-y}{1-p^2y}}}\int_{0}^{1}\frac{dx}{x-y}{\sqrt {\frac{x}{1-x}}}
\ln x
$$
$$
=-\frac{2{\sqrt b}}{\pi^2}p^2\int_{0}^{1}dy{\sqrt {\frac{1-y^2}{1-p^2y^2}}}
\frac{\partial}{\partial \nu}\int_{0}^{1}\frac{dx}{x-y^2}
\frac{x^{\nu-1}}{{\sqrt {1-x}}}\bigg|_{\nu=3/2}
$$
$$
=\frac{2{\sqrt b}}{\pi^2}p^2\int_{0}^{1}dy{\sqrt {\frac{1-y^2}{1-p^2y^2}}}
\frac{\partial}{\partial \nu}B(-1/2,\nu)F(3/2-\nu,1,3/2,1-y^2)
\bigg|_{\nu=3/2}.\eqno(B.2)
$$
We now focus on the derivative with respect to $\nu,$
$$
\frac{\partial}{\partial \nu}B(-1/2,\nu)
F(3/2-\nu,1,3/2,1-y^2)\bigg|_{\nu=3/2}
$$
$$
=\frac{\partial}{\partial \nu}B(-1/2,\nu)\bigg|_{\nu=3/2}F(0,1,3/2,1-y^2)
+B(-1/2,3/2)\frac{\partial}{\partial\nu}F(3/2-\nu,1,3/2,1-y^2)
\bigg|_{\nu=3/2}
$$
$$
=-2\pi(1-\ln 2)+\pi\frac{\partial}{\partial z}F(z,1,3/2,1-y^2)\bigg|_{z=0}
.\eqno(B.3)
$$
Note that we can make use of
$$\frac{\partial}{\partial z}\frac{\Gamma(z+a)}{\Gamma(z+b)}=
\frac{\Gamma(z+a)}{\Gamma(z+b)}\left[\psi(z+a)-\psi(z+b)\right],
$$
to compute the derivative of the Beta function with respect to its argument.
We now only have to compute the derivative of the Hypergeometric
function with respect to its first parameter.
As $z+1-3/2<0,$ for $z$ sufficiently
close to $0$, we may employ the power series representation of
$F(z,1,3/2,1-y^2)$ which is valid in $0\leq |1-y^2|<1$ and
differentiate term by term resulting in
$$
\frac{\partial}{\partial z}F(z,1,3/2,1-y^2)\bigg|_{z=0}
=\sum_{n=1}^{\infty}B(n,3/2)(1-y^2)^n
$$
$$
=\int_{0}^{1}\frac{dt}{t{\sqrt {1-t}}}\sum_{n=1}^{\infty}
\left[t(1-y^2)\right]^n
$$
$$
=(1-y^2)\int_{0}^{1}\frac{dt}{{\sqrt {1-t}}\left[1-t(1-y^2)\right]}
$$
$$
=(1-y^2)B(1,3/2)F(1,1,5/2,1-y^2)
$$
$$
=2-\frac{2y}{\sqrt {1-y^2}}\sin^{-1}({\sqrt {1-y^2}}).\eqno(B.4)
$$
{}From Eqs.(B.4),(B.3) and (B.2) we find
$$
I(a,b)=\frac{2{\sqrt b}}{\pi^2}
\left[-\frac{\pi^2}{p^2}+\int_{0}^{1}dy{\sqrt {\frac{1-y^2}{1-p^2y^2}}}
\left(\frac{2\pi(1-p^2)}{p^2(1-y^2)}+2\pi(1+\ln 2)\right)\right],\eqno(B.5)
$$
The integrals in Eq.(B.5) are recognized as the standard elliptic integrals
$(K\;{\rm and}\; D)$,
and we find,
$$
I(a,b)=2{\sqrt b}\left[-1+\left(\frac{2(1-p^2)}{\pi}
+\frac{2p^2(1+\ln 2)}{\pi}\right)K(p)-\frac{2p^2(1+\ln 2)}{\pi}D(p)\right]
.\eqno(B.6)
$$
Observe that $D(p)$ can be expressed in terms of $N$ and $K(p)$
[see Eq. (A.1)]. Further more using
$$p^2D(p)=K(p)-E(p),$$
where
$$
E(p)=\int_{0}^{1}dy{\sqrt {\frac{1-p^2y^2}{1-y^2}}},
$$
we have,
$$
I(a,b)=2\bt N\ln 2-2{\sqrt b}+\frac{4{\sqrt b}}{\pi}p^2E(p)
$$
and therefore
$$
A=\bt N(2\ln 2-\ln b-\ln p^2)+\frac{4{\sqrt b}}{\pi}p^2E(p).
$$
\newpage
\noindent
{\bf Appendix C Evaluation of the interaction energy}
\par\noindent
In this appendix we determine the interaction energy for $\ap>0,$
thus,
$$
\frac{1}{2}\int_{a}^{b}u(x)\sig(x)=I_1(a,b,\ap)+I_2(a,b,\ap),\eqno(C.1)
$$
where
$$
I_{1}(a,b,\ap):=-\frac{b^{2\ap-1}(b-a)}{2\pi^2\ap\bt}
\int_{0}^{1}dx\int_{0}^{1}\frac{dy}{x-y}
\frac{\left(\frac{1-p^2x}{1-p^2y}\right)^{\ap}-1}{y-x}
\left(1-p^2y\right)^{2\ap-1}{\sqrt {\frac{x}{y}}}{\sqrt {\frac{1-y}{1-x}}},
\eqno(C.2)$$
and
$$
I_2(a,b,\ap):=\frac{b^{2\ap-1}(b-a)}{2\pi^2\ap\bt}
\int_{0}^{1}dx\int_{0}^{1}\frac{dy}{x-y}(1-p^2y)^{2\ap-1}
{\sqrt {\frac{x}{y}}}{\sqrt {\frac{1-y}{1-x}}}.\eqno(C.3)
$$
We first compute $I_2(a,b,\ap)$,
$$
I_2(a,b,\ap)=\frac{a^{2\ap-1}(b-a)}{2\pi\ap\bt}\int_{0}^{1}dy(1-p^2y)^{2\ap-1}
{\sqrt {\frac{1-y}{y}}}=\frac{b-a}{2\bt}\cdot\frac{a^{2\ap-1}}{2\ap}
F(1-2\ap,1/2,2,p^2),
$$
and find
$$I_2(a,b,1/2)=\frac{b-a}{2\bt}.\eqno(C.4)$$
We now compute $I_1(a,b,\ap).$ Observe that,
$$
\frac{\left(\frac{1-p^2x}{1-p^2y}\right)^{\ap}-1}{y-x}
$$
$$
=\frac{\left[1-\frac{p^2(x-y)}{1-p^2y}\right]^{\ap}-1}{y-x}
$$
$$
=-\int_{0}^{\frac{p^2}{1-p^2y}}\frac{d\epsilon}{x-y}
\frac{\partial}{\partial \epsilon}\left[1-\epsilon(x-y)\right]^{\ap}
$$
$$
=\ap\int_{0}^{\frac{p^2}{1-p^y}}d\epsilon(1+\epsilon y)^{\ap-1}
\left(1-\frac{\epsilon x}{1+\epsilon y}\right)^{\ap-1}.
$$
With the change of variable
$$
\frac{\epsilon}{1+\epsilon y}=t,$$
we get
$$
\frac{\left(\frac{1-p^2x}{1-p^2y}\right)^{\ap}-1}{y-x}
=\ap\int_{0}^{p^2}\frac{dt}{(1-ty)^{1+\ap}(1-tx)^{1-\ap}}.\eqno(C.5)
$$
Using Eq. (C.5) in Eq. (C.2),
$$
I_1(a,b,\ap)=-\frac{b^{2\ap-1}(b-a)}{2\pi^2\bt}
\int_{0}^{1}dx{\sqrt {\frac{x}{1-x}}}
\int_{0}^{1}dy{\sqrt {\frac{1-y}{y}}}\int_{0}^{p^2}\frac{dt}{(1-ty)^{1+\ap}
(1-tx)^{1-\ap}}
$$
$$
=-\frac{b^{2\ap-1}(b-a)}{2\pi^2\bt}B(1/2,3/2)
\int_{0}^{p^2}dtF(1-\ap,3/2,2,t)\int_{0}^{1}dy\frac{(1-p^2y)^{2\ap-1}}
{{1-ty}^{1+\ap}}{\sqrt {\frac{1-y}{y}}}.\eqno(C.6)
$$
For $\ap=1/2$, Eq. (C.6) becomes
$$
I_1(a,b,1/2)=-\frac{b-a}{2\pi^2\bt}B^2(1/2,3/2)\int_{0}^{p^2}dtF(1/2,3/2,2,t)
F(3/2,1/2,2,t)
$$
$$
=-\frac{b-a}{8\bt}\int_{0}^{p^2}dt\left[D\left({\sqrt t}\right)\right]^2,
\eqno(C.7)
$$
where we have used
$F(3/2,1/2,2,t)=F(1/2,3/2,2,t)=\frac{4}{\pi}D\left({\sqrt t}\right),$
to get Eq. (C.7). From Eqs. (C.4) and (C.7), the interaction energy
reads
$$
I_1(a,b,1/2)+I_2(a,b,1/2)
=\frac{b-a}{2\bt}\left(1-\frac{4}{\pi^2}\int_{0}^{p^2}
dt\left[D\left({\sqrt t}\right)\right]^2\right).\eqno(C.8)$$
Note that
$$
\int_{0}^{p^2}dt\left[D\left({\sqrt t}\right)\right]^2
=\int_{0}^{1}dt\left[D\left({\sqrt t}\right)\right]^2
-\int_{0}^{a/b}dt\left[D\left({\sqrt t}\right)\right]^2,\eqno(C.9)
$$
since $p^2=1-a/b.$ With the aid of
$$\int_{0}^{1}dt\left[D\left({\sqrt t}\right)\right]^2
=\frac{\pi^2}{4}-1,$$
and Eq. (C.9), we arrive at Eq. (60) in the text.


\begin{thebibliography}{99}
\bibitem{Mourad} We would like thank Mourad Ismail for discussions.
\bibitem{Chen7} For the Gaussian Ensembles the scaling variable should
be $\sim N^{1/6},$ according to our formula.
\end{thebibliography}
\end{document}